\documentclass[prl,aps,twocolumn,showpacs,superscriptaddress,amsfonts,amsmath,floatfix]{revtex4}

\usepackage{graphicx}
\usepackage{color}
\usepackage{import}
\usepackage{wrapfig}
\usepackage{mathtools}
\usepackage{bm}
\usepackage{amsmath}
\usepackage{amsfonts}
\usepackage{amssymb}

\newcommand*\conj[1]{\overline{#1}}

\begin{document}


\title{Synthetic anyons in noninteracting systems}

\author{Frane Luni\'{c}}
\affiliation{Department of Physics, Faculty of Science, University of Zagreb, Bijeni\v{c}ka c. 32, 10000 Zagreb, Croatia}

\author{Marija Todori\'{c}}
\affiliation{Department of Physics, Faculty of Science, University of Zagreb, Bijeni\v{c}ka c. 32, 10000 Zagreb, Croatia}

\author{Bruno Klajn}
\affiliation{Department of Physics, Faculty of Science, University of Zagreb, Bijeni\v{c}ka c. 32, 10000 Zagreb, Croatia}

\author{Tena Dub\v{c}ek}
\affiliation{Institute for Theoretical Physics, ETH Z\"urich, 8093 Zurich, Switzerland}

\author{Dario Juki\'{c}}
\affiliation{Faculty of Civil Engineering, University of Zagreb, A. Ka\v{c}i\'{c}a  Mio\v{s}i\'{c}a 26, 10000 Zagreb, Croatia}

\author{Hrvoje Buljan}
\email{hbuljan@phy.hr}
\affiliation{Department of Physics, Faculty of Science, University of Zagreb, Bijeni\v{c}ka c. 32, 10000 Zagreb, Croatia}
\affiliation{The MOE Key Laboratory of Weak-Light Nonlinear Photonics, 
TEDA Applied Physics Institute and School of Physics, Nankai University, Tianjin 300457, China}

\date{\today}

\begin{abstract}
Synthetic anyons can be implemented in a noninteracting many-body system, by using specially tailored localized (physical) probes, which supply the demanded nontrivial topology in the system. We consider the Hamiltonian for noninteracting electrons in two-dimensions (2D), in a uniform magnetic field, where the probes are external solenoids with a magnetic flux that is a fraction of the flux quantum. 
The Hamiltonian could also be implemented in an ultracold (fermionic) atomic gas in 2D, in a uniform synthetic magnetic field, where the probes are lasers giving rise to synthetic solenoid gauge potentials. 
We find analytically and numerically the ground state of this system when only the lowest Landau level states are occupied.
It is shown that the ground state is anyonic in the coordinates of the probes. 
We show that these synthetic anyons cannot be considered as emergent quasiparticles. 
The fusion rules of synthetic anyons are discussed for different microscopic realizations of the fusion process. 
\end{abstract}

\maketitle

\section{Introduction}

Anyons are quantum particles that exist in two-dimensional (2D) space~\cite{Wilczek1982, Leinaas1977}. 
Their exchange statistics interpolates between bosons and fermions, which gives rise to intriguing and 
nontrivial quantum mechanical properties of anyonic systems~\cite{Wilczek1982}. 
The Fractional Quantum Hall Effect (FQHE)~\cite{Tsui1982, Laughlin1983} is a paradigm of anyonic systems; 
emergent quasiparticles upon FQHE state(s) behave as anyons~\cite{Arovas1984, Camino2005}. 
More recent examples include spin systems~\cite{Kitaev2003, Kitaev2006, Dai2017, Klanjsek2018} and Majorana zero modes~\cite{DasSarma2015, Mourik2012}. 
The so-called non-Abelian anyons were proposed to lead to topologically protected quantum computing~\cite{Kitaev2003, Nayak2008}. 
However, there is still a long way to go before experiments will be able to efficiently detect and manipulate 
anyons, especially for fault tolerant quantum computing~\cite{Nayak2008, DasSarma2015}. 
Thus, there is a motivation to explore less traditional schemes for realizing and manipulating anyons. 
For example, it was proposed that anyons could be synthesized by coupling weakly interacting (or noninteracting) electrons to a
topologically nontrivial background (or topologically nontrivial external perturbations)~\cite{Weeks2007, Rosenberg2009, Rahmani2013}.

We investigate the potential implementation of such synthetic anyons by studying a Hamiltonian for noninteracting 2DEG, in a uniform magnetic field, with $N$ external solenoids (probes), which introduce localized fluxes at positions $\pmb{\eta}_k$, $k=1,\ldots,N$.
The Hamiltonian could in principle also be implemented in an ultracold (fermionic) atomic gas in 2D, in a uniform synthetic gauge field, where the probes are lasers which give rise to synthetic solenoid gauge potentials. 
We find analytically and numerically the ground state of this Hamiltonian when the 
Fermi energy is such that only the lowest Landau level states are populated.
When the flux through a solenoid $\Phi$ is a fraction of the flux quantum, $\Phi=\alpha\Phi_0$, the ground state 
wavefunction is anyonic in the coordinates of the external probes $\pmb{\eta}_k$. 
In other words, by braiding the probes one imprints the Berry (statistical) phase~\cite{Berry1984} on the ground state.  
Around every solenoid probe there is a cusp-like dip of missing electron charge $\Delta q$. 
We demonstrate that the missing charge should not be identified with the concept of an emergent quasiparticle by 
showing that $\frac{\Delta q}{\hbar}\oint \mathbf{A}\cdot d {\mathbf l} $ does not correspond to the Aharonov-Bohm phase~\cite{Aharonov1959} acquired as the probe traverses a loop in space.
One could arrive at the same conclusion by using gauge invariance arguments~\cite{Rahmani2013}.
This has consequences on the fusion rules of these synthetic anyons: the fusion rules depend 
on the microscopic details of the fusion process as discussed below. 
Even though we consider Abelian anyons, if an analogous scheme for synthetic non-Abelian anyons is developed, it will be a potential path towards a platform for quantum computers, which further motivates this study.

In addition to the condensed matter experiments on the FQHE~\cite{Tsui1982, Camino2005}, Majorana zero modes~\cite{Mourik2012}, and anyons in the Kitaev paramagnetic state of the honeycomb magnet α-RuCl3~\cite{Klanjsek2018} (see Refs.~\cite{Nayak2008, DasSarma2015} for reviews), anyonic behavior was experimentally addressed in other systems. 
The Kitaev toric model~\cite{Kitaev2003} was concieved as a platform for topological quantum computing employing non-Abelian anyons. 
Its minimal variant was experimentally realized in ultracold atomic gases~\cite{Dai2017}, and with trapped ions using dissipative
pumping processes~\cite{Barreiro2011}. Anyonic statistics was simulated in photonic quantum simulators~\cite{Lu2009, Pachos2009}, 
superconducting quantum circuits~\cite{Zhong2016}, and by using nuclear magnetic resonance~\cite{Li2017}. 
The body of theoretical proposals is larger (we will not attempt to provide a review), and 
besides the condensed matter systems~\cite{Nayak2008, DasSarma2015}, 
includes proposals in ultracold atomic gases based on emulating the FQHE~\cite{Paredes2001, Zhang2014}, 
the Kitaev model~\cite{Duan2003, Jiang2008}, or by employing synthetic gauge potentials~\cite{Burrello2010}.
Moreover, different mechanisms to achieve FQH states of light have been proposed~\cite{Kapit2014, Umucalilar2017}.

The aforementioned less traditional schemes for realizing anyons include a system of an artificially structured type-II superconducting
film, adjacent to a two-dimensional electron gas (2DEG) in the {\em integer} QHE with unit filling fraction~\cite{Weeks2007,Rosenberg2009}, anyons in {\em integer} QHE magnets~\cite{Rahmani2013}, 
and topological defects in graphene~\cite{SeradjehPRL2008}, where fractional statistics appears for weakly 
interacting electrons coupled to a topologically nontrivial background (e.g. vortex) or to external perturbations. 
It was recently proposed that a charge-flux composite (i.e., anyon) can be achieved by sandwiching a charged magnetic dipole 
between two semi-infinite blocks of a high permeability material~\cite{Todoric2018}. 
Here we study the possibility to synthesize anyons upon a noninteracting many-body ground state by employing specially tailored localized probes.

\section{The ground state wavefunction}

\begin{figure}
\includegraphics[width=0.46\textwidth]{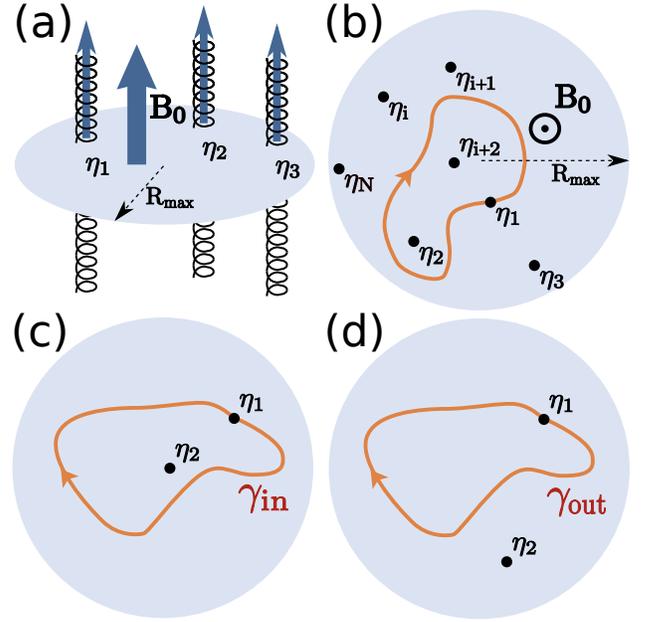}
\caption{Sketch of the system. (a) We explore a 2DEG in a magnetic field ${\bf B_0}$, on a disc of radius $R_{max}$. The solenoid probes with flux $\Phi$, pierce the 2DEG at positions $\eta_j$ (coordinates are in complex notation). (b) The contour path of one probe, which adiabatically traverses a closed loop in space; we are interested in the Berry phase accumulated along such paths. Illustration of the contours corresponding to $\gamma_{in}$ (c), and $\gamma_{out}$ (d). See text for details. }
\label{fig:sketch}
\end{figure}

We consider $N_e$ noninteracting spin-polarized electrons in 2D configuration space (in the $xy$ plane), in a uniform magnetic field ${\bf B}_0=\bm{\nabla}\times {\bf A}_0=B_0 \bm{\hat z}$, where ${\bf A}_0({\bf r})={\bf B}_0\times {\bf r}/2$ is the vector potential in the symmetric gauge ($B_0>0$). The system is perturbed with $N$ very thin solenoids at locations $\pmb{\eta}_k=\eta_{x,k}\bm{\hat x}+\eta_{y,k}\bm{\hat y}$. The vector potential of each solenoid is  
\begin{equation}
{\bf A}_k({\bf r})=\frac{\Phi}{2\pi} \frac{\bm{\hat{z}}\times({\bf r}-\pmb{\eta}_k)}{|{\bf r}-\pmb{\eta}_k|^2},
\end{equation}
where $\Phi$ is the magnetic flux through a solenoid. 
The Hamiltonian of this system is then 
\begin{equation}
 H=\sum_{j=1}^{N_e}\frac{1}{2m}\left({\bf p}_j-q{\bf A}_0({\bf r}_j)-q\sum_{k=1}^{N}{\bf A}_k({\bf r}_j) \right)^2+\sum_{j=1}^{N_e} V({\bf r}_j),
\label{Ham}
\end{equation}
where $V({\bf r})$ is zero for $r<R_{max}$, and infinite otherwise; $q<0$ ($m$) is the electron charge (mass, respectively).  The system is illustrated in Fig. \ref{fig:sketch}(a). We assume that the Fermi level is such that only the states from the lowest Landau level (LLL) of energy $\hbar \omega_B/2$ are populated ($\omega_B=-qB_0/m$), and we assume they are {\em all} populated. The many-body ground state of this system is denoted by $\psi(\{ z_j \},\{\bar z_j\};\{ \eta_k \},\{\bar \eta_k\})$, where $z_j=x_j+iy_j$ and $\bar z_j=x_j-iy_j$ are the electron coordinates, and $\eta_k=\eta_{x,k}+i\eta_{y,k}$ and $\bar \eta_k=\eta_{x,k}-i\eta_{y,k}$ are the probe coordinates in complex notation.

In this section we demonstrate that the ground state wavefunction with energy $N_e \hbar \omega_B/2$ is given by 
\begin{eqnarray}
\psi  & = & \frac{1}{\sqrt{Z(\{ \eta_k \},\{\bar \eta_k\})}} \left[ \prod_{j=1}^{N_e} \prod_{k=1}^N
|z_j-\eta_k|^{-\alpha}\overline{z_j-\eta_k} \right]  \nonumber \\
& & \times \left[ \prod_{i<j}^{N_e}(\bar{z}_i-\bar{z}_j) \right]
\exp\left(-\sum_{i=1}^{N_e} \frac{|z_i|^2}{4l_B^2}\right),
\label{psiNprobes}
\end{eqnarray}
where $l_B=\sqrt{-\hbar/B_0 q}$ is the magnetic length, $\alpha=\Phi/\Phi_0$, $\Phi_0=-2\pi\hbar/q$ is the flux quantum, and $Z(\{ \eta_k \},\{\bar \eta_k\})$ accounts for normalization. For clarity, we consider $\alpha \in (0,1)$; 
results for fractional values outside of the $(0,1)$ interval are easily deduced.

\begin{figure}
\includegraphics[width=0.46\textwidth]{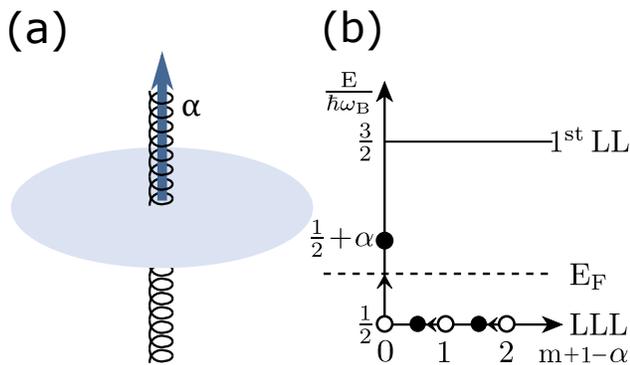}
\caption{Sketch of the energy scales and the spectral flow for just one probe. (a) A probe is centered in the system, its flux is such that $0\leq\alpha=\Phi/\Phi_0\leq 1$. (b) As $\alpha$ is increased, there is a spectral flow as illustrated. The Fermi energy $E_F$ is always set such that only the LLL states are populated. See text for details. }
\label{fig:spectral_flow}
\end{figure}

For the clarity of the presentation, we first present what happens with the system when only one probe is placed in the system, and subsequently what happens when two probes are inserted. For a single probe, the single-particle states of the system at the LLL energy are given by (see Appendix for details of the calculation)
\begin{equation}
\psi_m=|z-\eta|^{-\alpha}\, \conj{z-\eta}\, \bar{z}^m \exp\left(-\frac{|z|^2}{4l_B^2}\right), \ m=0,1,2,\ldots.
\label{wf}
\end{equation}
There is one state localized at the position of the probe, with energy $\hbar\omega_B(1+2\alpha)/2$ in between LLL and the first excited Landau level:
\begin{equation}
\psi_{LS}=|z-\eta|^\alpha  \exp\left(-\frac{|z-\eta|^2 + \bar{\eta} z - \eta \bar{z}}{4l_B^2}\right).
\label{excited}
\end{equation}
Suppose that one introduces the solenoid probe at some point in time, and adiabatically increases the flux through it. As $\alpha$ increases from zero to one, there is spectral flow illustrated in Fig. \ref{fig:spectral_flow}; one state from the LLL rises in energy and flows towards the first Landau level. When $\alpha=1$, this flux is just gauge, and the energies map back onto those at $\alpha=0$. This scenario is well known from studies of the QHE~\cite{Tong}. Here we assume that the flux is fixed at some value $\alpha$, and the Fermi energy is between the LLL energy and $\hbar\omega_B(1+2\alpha)/2$; thus, this localized state is not populated. The many-body ground state is constructed by inserting all LLL states in a Slater determinant; it is given by Eq. (\ref{psiNprobes}) for $N=1$.

For the case of two probes, the single-particle states of the system at the LLL energy are 
\begin{eqnarray}
\psi_m & = & |z-\eta_1|^{-\alpha}\, |z-\eta_2|^{-\alpha}\, \conj{z-\eta_1}\, \conj{z-\eta_2} \nonumber \\
 & & \times \bar{z}^m\, \exp \left(-\frac{|z|^2}{4l_B^2}\right) , \  m=0,1,2,\dots
\label{twosol}
\end{eqnarray}
Now there are two localized states in between the LLL and the first excited Landau level. We did not find analytical expressions for these states, but they are visible in numerical calculations. The energies of these localized states are in the gap, between the LLL and the first excited Landau level. They increase with increasing alpha and join the first excited Landau level when $\alpha=1$ as expected. The many-body ground state is given by Eq. (\ref{psiNprobes}) for $N=2$. 

\begin{figure}
\includegraphics[width=0.45\textwidth]{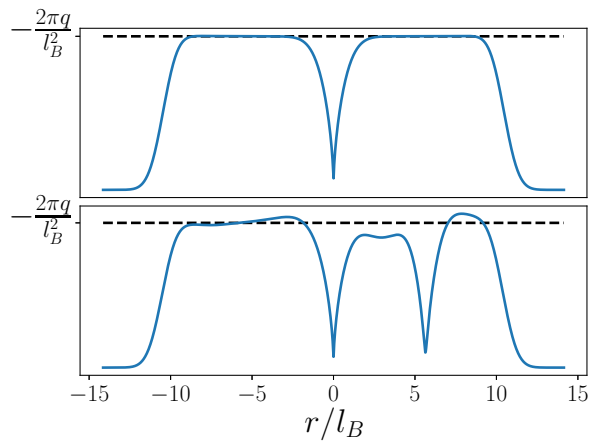}
\caption{The single-particle densities (cross sections) of the ground states with one 
probe (at $r=0$), and and two probes (at $r=0$ and $r=5.264 l_B$). 
The flux through the probes is given by $\alpha=0.7$. 
Horizontal dashed line depicts the density of an infinite system, see text for details. }
\label{fig:SPdensity}
\end{figure}

Now we generalize our results for any number of the probes $N$. 
To this end, we employ the following singular gauge transformation:
\begin{equation}
\psi'=\psi \prod_{i=1}^{N_e} \prod_{j=1}^{N} \exp(i \alpha \phi_{ij});
\label{multival}
\end{equation}
here $\phi_{ij}$ denotes the argument of $z_i-\eta_j=|z_i-\eta_j|\exp(i\phi_{ij})$. In this gauge, the vector potential of the probes is ${\bf A}'_k=\bm{0}$ everywhere except at the positions of the probes, and the Hamiltonian $H'$ is given by Eq. (\ref{Ham}) with ${\bf A}_k$ replaced by ${\bf A}'_k=\bm{0}$. It is straightforward to verify that $\psi'$ is an eigenstate of $H'$ with energy $N_e\hbar\omega_B/2$, and hence the ground state.

In Figure \ref{fig:SPdensity} we illustrate the single-particle density (cross section) for the system with one and two probes. 
Clearly, the single-particle density has a cusp-like dip at the position of a probe, i.e., a missing electron charge $\Delta q>0$. 
It is tempting to identify the composite of the missing electron charge $\Delta q$ and the probe with flux $\Phi$ with 
Wilczek's charge-flux-composite anyons~\cite{Wilczek1982}; however, a careful analysis of the Berry phase 
below shows that this identification would be erroneous.

In a potential experimental implementation of the proposed system, one should not populate the localized states such as $\psi_{LS}$. With this state populated, the ground state is no longer anyonic in the coordinates of the probes. 
For this state to remain empty, the temperature must be sufficiently low such that $kT \ll \hbar \omega_B \alpha$, which is difficult to obtain for small $\alpha$. However, an additional localized repulsive scalar potential at location of the probes (e.g., delta function potential), which may be present naturally depending on the realization, would lift the energies of the localized states to remedy this issue.

To end this section, let us mention that when calculating the single-particle states of the LLL, which enter the Slater determinant used to construct the ground state in Eq.~(\ref{psiNprobes}), one encounters a spurious single-particle state of the form
\begin{equation}
\psi_\text{spur} = |z|^{-\alpha} \exp \left(-\frac{|z|^2}{4l_B^2}\right),
\label{spurious}
\end{equation}
which, although normalizable, has divergent density. The form (\ref{spurious}) corresponds to a system with a single probe centered at the origin. A more careful analysis (see Appendix for details) shows that this state is, in fact, not an eigenstate of the Hamiltonian and should not be used in the construction of the Slater determinant. If this state was physical and present in the ground state, the ground state would not be anyonic in the coordinates of the probes. In that case however, the aforementioned additional localized repulsive scalar potential at location of the probes could be used to lift it in energy and remove it from the ground state. We should note that in Ref. \cite{Weeks2007} this spurious state was used to construct the many-body ground state, and as a result the ground state from Ref. \cite{Weeks2007} is in fact not anyonic (see below our discussion on gauge invariance in calculating the Berry phase).

\section{Anyonic properties of the wavefunction - calculation of the Berry phase}

In this section we calculate the Berry phase as one of the probes undergoes adiabatically a closed loop in space as illustrated in Fig. \ref{fig:sketch}(b). The Berry phase depends on how many other probes are contained in the loop. More specifically, following a similar calculation as Arovas et al.~\cite{Arovas1984}, we calculate the Berry phase when a single probe is within the loop (call it $\gamma_{in}$, see Fig. \ref{fig:sketch}(c)), and when all of the other probes are outside of the loop (call it $\gamma_{out}$, see Fig. \ref{fig:sketch}(d)). The difference between the two phases is the statistical phase, which we find to be $\gamma_S=\gamma_{in}-\gamma_{out}=2\pi(\alpha-1)$, where $\alpha=\Phi/\Phi_0$. This result means that in the coordinates of the external probes, the wavefunction $\psi$ is anyonic when $\alpha$ is fractional. 

We outline the derivation below, while details of the calculation are in the Appendix. 
We assume that the probes remain sufficiently far apart from each other at any time.  
Without loosing any generality, we assume that the probe $\eta_1$ traverses the path. 
The Berry phase $\gamma$ accumulated along the path $C$ is given by
\begin{equation}
 \gamma=-\oint_C \left [ \mathcal{A}_{\eta_1}d \eta_1 + \mathcal{A}_{\bar{\eta}_1}d\bar{\eta}_1 \right ].
\label{Berry}
\end{equation}
The holomorphic Berry connection is given by
\begin{equation*}
    \mathcal{A}_{\eta}(\eta,\bar{\eta})=-i\langle \psi|\frac{\partial}{\partial \eta}|\psi\rangle,
\end{equation*}
while the anti-holomorphic Berry connection is 
\begin{equation*}
    \mathcal{A}_{\bar{\eta}}(\eta,\bar{\eta})=-i\langle\psi|\frac{\partial}{\partial \bar{\eta}}|\psi\rangle.
\end{equation*}
By following Ref.~\cite{Arovas1984}, and in addition by taking the normalization $Z$ into account by employing the plasma analogy~\cite{Laughlin1983,Tong}, we find that the Berry phase accumulated in this process is
\begin{equation}
\gamma=2\pi \langle n \rangle_C,
\end{equation}
where $\langle n \rangle_C$ is the mean number of electrons in the area encircled by the path $C$. 
It is evident that $\gamma_{in}$ will differ from $\gamma_{out}$. 
In the former case, the inner probe expels some charge away from itself as illustrated in Fig. \ref{fig:SPdensity}, 
and the mean number of electrons inside the contour differs in the two cases: $\langle n \rangle_{C,in}\neq \langle n \rangle_{C,out}$. 
The difference between these two cases is the statistical phase 
\begin{equation}
\gamma_S=2\pi (\langle n \rangle_{C,in}-\langle n \rangle_{C,out}). 
\end{equation}
We calculate the expelled charge from the single-particle density $\rho$ of the many-body wavefunction $\psi$. 
It can be found by employing the plasma analogy~\cite{Laughlin1983,Tong}:
\begin{equation*}
\rho(z)=\frac{1}{2\pi l_B^2}-(1-\alpha)\sum_{k=1}^N\delta^2(z-\eta_k).
\end{equation*}
The missing charge at the probe is thus $\Delta q=-q(1-\alpha)$, and 
the statistical phase (in the thermodynamic limit) is 
\begin{equation}
\gamma_S=2\pi (\alpha-1).
\label{statphase}
\end{equation}
Thus, $\gamma_S \text{ mod } 2\pi$ is equal to $2\pi\alpha$.

In order to further underpin this result, and explore the dependence of the statistical phase on the separation between the probes (we approximated above that they are sufficiently far apart along the path $C$) and the details of the path, we perform numerical calculations. 
We numerically consider the cases with one and two probes. In all our calculations presented here, the magnetic field is given by $B_0 R_\mathrm{max}^2\pi / \Phi_0 = 100$; we construct the numerical ground state by filling the first $N_{e}=55$ states to minimize the boundary (finite size) effects, and still successfully mimic an infinite system.
The method for the numerical calculation of the Berry phase is as follows~\cite{Mukunda1993}: instead of performing the integral in \eqref{Berry}, we discretize the evolution parameter, here called time for simplicity, and evaluate it at $T$ equidistant points. Let $\psi_i(t)$ be the $i$-th numerical single-particle eigenstate lying in the LLL at time $t$, and let $M_{ij}(t_k,t_\ell) = \langle \psi_i(t_k)| \psi_j(t_\ell)\rangle$ be the elements of the overlap matrix $M(t_k,t_\ell)$ at two different times. Then the Berry matrix
\begin{equation}\label{eq:berry_matrix}
U = M(t_0,t_1) M(t_1,t_2)\dots M(t_T,t_0)
\end{equation}
leads directly to the Berry phase
\begin{equation}
\gamma \approx -\operatorname{arg}(\det U).
\label{eq:berry_phase}
\end{equation}
This relation is exact in the limit $T\to \infty$. As before, the statistical phase is $\gamma_{S}=\gamma_{in}-\gamma_{out}$.

\begin{figure}
\includegraphics[width=0.45\textwidth]{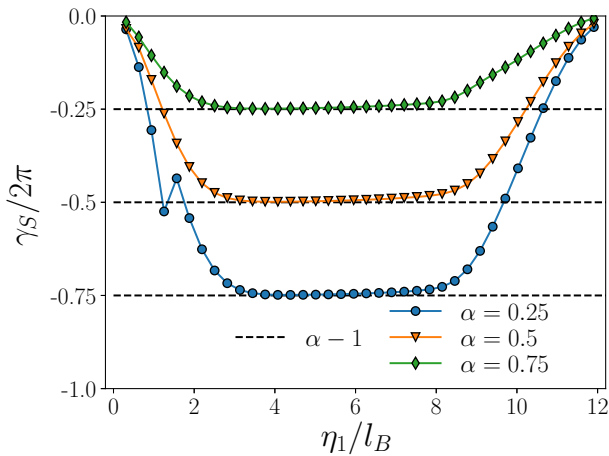}
\caption{The statistical phase $\gamma_{S}$ as a function of the separation between the probes at three different flux values $\alpha$. In every calculation, one of the probes is at $z=0$, and the other one adiabatically traverses a circle of radius $R$ in the counterclockwise direction. The dashed lines denote the $2\pi(\alpha-1)$ values corresponding to the analytical prediction. 
}
\label{fig:statph}
\end{figure}

In Fig. \ref{fig:statph} we illustrate $\gamma_{S}$ as a function of the separation between the probes $R$. The dashed lines denote the analytical prediction Eq. (\ref{statphase}). We see that if the probes are too close, they will influence each others cusp-dip in the density, and consequently the statistical phase will not be given by $2 \pi (\alpha-1)$. However, after they are sufficiently apart, $\gamma_{S}$ exhibits a plateau at the value of $2 \pi (\alpha-1)$. 
As the outer probe gets close to the edge of our (numerical) finite size system, the phase departs from the analytical solution. 
We conclude that the numerical calculations agree with the analytical prediction when the path of the moving probe does not come too close to other probes, and if they are not too close to the edges of the sample. The system exploited in numerical calculations is very small (practically mesoscopic). In reality, the system would be much larger providing a much broader region in space where a constant plateau would be observed.

\begin{figure}
\includegraphics[width=0.45\textwidth]{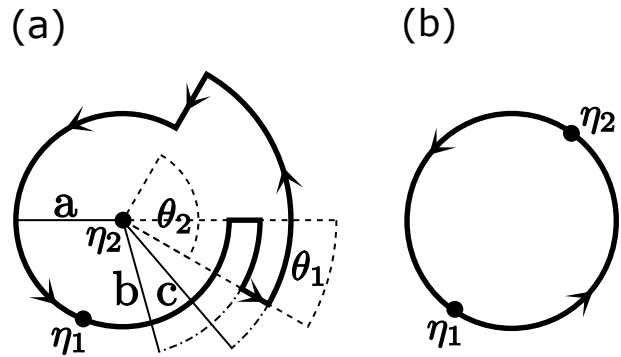}
\caption{Two different contours. (a) One of the probes undergoes a closed loop, visiting three different radii $a$, $b$, and $c$, such that each is sufficiently far from the probe at zero and from the edge of the system. (b) Two probes at opposite radii ($|\eta_1|=|\eta_2|=3.13l_B$) are exchanged leading to a phase $\pi(\alpha-1)$. The parameters used in the calculation are $\alpha = 0.37$, $a=4.76l_B$, $b=6.14l_B$, $c=7.52l_B$, $\theta_1=\pi/6$, and $\theta_2=\pi/2$.}
\label{fig:contours}
\end{figure}

Next we perform the same calculation, but deform the contour $C$ as illustrated in Fig. \ref{fig:contours}(a). 
The contour is such that the probes are sufficiently separated at all times, and away from the sample edges. 
We obtain the statistical phase $\gamma_{S}=-0.631\times 2\pi$, which is in agreement with the analytical result 
$\gamma_{S}=2\pi(\alpha-1)$, with relative error of about $0.2\%$. 
We conclude that the statistical phase does not depend on the details of the contour.

Next we consider the exchange of two probes. We numerically calculate the phase obtained when two of the 
probes are exchanged along the path illustrated in Fig. \ref{fig:contours}(b). 
For the exchange phase corresponding to the exchange path in Fig. \ref{fig:contours}(b) we obtain $-0.636\times \pi$, once again in agreement with analytical calculations. The relative error is about $1\%$.
From the viewpoint of the relative coordinate, when one of the probes encircles the other probe, 
this corresponds to a double exchange of the two probes illustrated in Fig. \ref{fig:contours}(b). 
Thus, we conclude that if we exchange two of the probes adiabatically along a path illustrated in Fig. \ref{fig:contours}(b) (with no other probes within the closed contour), the exchange phase accumulated by the wavefunction will be $\pi(\alpha-1)$. This means that the wavefunction $\psi$ is anyonic in the coordinates of the probes, with the statistical parameter given by $\theta = \pi(\alpha-1)$.

We end this section by a note on the gauge invariance of the Berry phase calculated along the closed path $C$. 
The wavefunction $\psi$ in Eq. (\ref{psiNprobes}) is a single-valued function of the positions of the external probes $\eta_k$, provided that the normalization $Z(\{ \eta_k \},\{\bar \eta_k\})$ is also chosen to be a single-valued function of $\eta_k$. In contrast, the singular gauge wavefunction $\psi'$ in Eq. (\ref{multival}) is a multivalued function of $\eta_k$.
Equation (\ref{Berry}) for calculating the Berry phase yields different results when naively used for $\psi$ and $\psi'$. 
However, the Berry phase calculated along a closed path must be independent of the gauge used. 
This issue is resolved by noting that Eq. (\ref{Berry}) should be used only for single-valued wavefunctions (that is $\psi$ in our case). 
If one wishes to calculate the Berry phase in the singular gauge by using the multivalued wavefunction $\psi'$, 
there is an additional term that should be included in the Berry phase formula (see Eq. (5.12) in \cite{Mukunda1993}) which ensures gauge invariance. 
We note that our results differ from Refs. \cite{Weeks2007, Rosenberg2009}, which have used multivalued wavefunctions and 
Eq. (\ref{Berry}) to calculate the Berry phase. 

\section{Synthetic anyons are not emergent quasiparticles}

From the illustration of the single-particle density in Fig. \ref{fig:SPdensity} we see that at the position of every 
solenoid probe there is a cusp-like dip, i.e., a missing electron charge, which is found to be 
$\Delta q=-q(1-\alpha)$ from the single-particle density. 
We have already noted that it is tempting to identify the composite of a missing electron charge $\Delta q$, 
and a solenoid with flux $\Phi$ with Wilczek's charge-flux-composite anyon~\cite{Wilczek1982}. 
Now we show that such an interpretation is erroneous.

When a probe traverses a closed path $C$, the system acquires the Berry phase $\gamma=2\pi \langle n \rangle_C$. 
Let us try to calculate the missing charge by a different route using the Aharonov-Bohm phase, 
and by assuming that we are dealing with a charge-flux-composite. 
To this end, let us denote the missing charge $q^*$, and check whether we obtain the same result as 
with the single-particle density. When the charge $q^*$ traverses the path $C$, it will acquire the 
Aharonov-Bohm phase $q^{*}\Phi_{C}/\hbar$, where $\Phi_{C}=\langle n \rangle_C\Phi_0$ is the total magnetic flux within 
the path $C$ (we have assumed unity filling of the LLL). To obtain the Berry phase, we should include the Aharonov-Bohm 
phase acquired by the solenoid with flux $\alpha\Phi_0$ that circulates around the charge $q\langle n \rangle_C$, which is equal to 
$q \langle n \rangle_C \alpha \Phi_0/\hbar$. By identifying 
\begin{equation*}
\gamma=2\pi \langle n \rangle_C=\frac{q^{*}\Phi_{C}}{\hbar}
+
\frac{q\langle n \rangle_C \alpha \Phi_0}{\hbar},
\end{equation*}
we find 
\begin{equation*}
q^{*}=-q(1+\alpha)\neq \Delta q=-q(1-\alpha).
\end{equation*}
This difference may seem as a surprise, because an equivalent calculation for 
anyons in the FQHE yields identical expressions for the missing charge from 
the single-particle density and from the Aharonov-Bohm calculation of $q^{*}$. 

To understand the obtained result, first we note that the external solenoid probe acts as a ladle that 
steers the electron see around, and the Aharonov-Bohm phase depends on the movements 
of the electrons in the see, and not of the missing charge. 
When the missing charge corresponds to the quasiparticle, as in the 
FQHE, then $q^{*}=\Delta q$ because the motion of (quasi)holes 
uniquely corresponds to the motion of the electron sea. 
However, the missing charge here is not a quasihole, and we cannot interpret 
the missing charge attached to the solenoid probe as Wilczek's charge-flux-tube composite. 
One way to understand this difference is to assume that the electron sea 
is a superfluid, the Aharonov-Bohm phase acquired by steering the ladle would be zero.

\section{Fusion rules of synthetic anyons}

The conclusion of the previous Section has impact on the fusion rules of synthetic anyons. 
The fusion rules depend on the physical microscopic process which corresponds to the fusion. 
For example, suppose that we have $N=4$ solenoid probes in the system with flux 
$\alpha\Phi_0$, i.e., we have two pairs of probes. Next, we slowly bring together (merge) two of the solenoids
from each pair, thereby forming a system with $N=2$ solenoid probes with flux 
$2\alpha\Phi_0$. This system is identical to the one we have explored with $\alpha$ replaced 
with $2\alpha\, \mbox {mod}\, 1$. Thus, the exchange phase changes from $\pi (\alpha-1)$
to $\pi((2\alpha\,  \mbox{mod}\, 1) -1)$. This is not the exchange phase $2^2\pi (\alpha-1)$ expected 
from fusing two anyons. This is related to the fact that we cannot interpret 
the missing charge attached to a solenoid probe as Wilczek's charge-flux-tube composite,
because in that case the standard fusion rules would be applicable.

If we, however, consider the fusion process as pairing the solenoids two-by-two in the sense 
$\eta_2=\eta_1+c$, and $\eta_4=\eta_3+c$, where $c$ is a complex number with magnitude greater than 
$l_B$, then the standard fusion rules apply. For example, if we move one of the pairs in a circle of sufficiently large radius around the other pair, we analytically obtain the expected statistical phase of $2^2\times 2\pi (\alpha-1)$ (see Appendix for details of calculation).

\begin{figure}
\includegraphics[width=.40\textwidth]{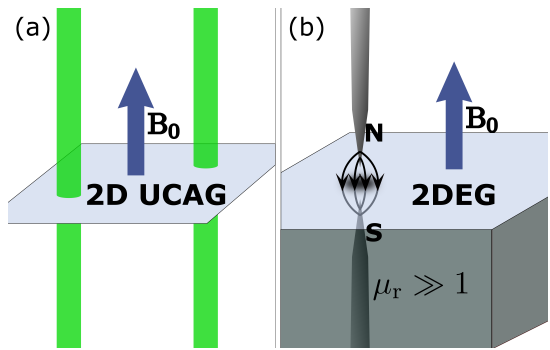}
\caption{Sketch of the potential experimental implementations of the studied Hamiltonian. 
(a) Two-dimensional ultracold atomic gas in a synthetic homogeneous magnetic field, perturbed with laser beams giving rise to solenoid gauge potentials. 
(b) A magnetic needle suspended above a 2DEG heterostructure grown on a material with high magnetic permeability $\mu_r\gg 1$. See text for details.}
\label{fig:experiment}
\end{figure}

\section{Potential experimental implementation of the Hamiltonian}

Next we discuss the potential implementations of Hamiltonian (\ref{Ham}) in experiments. 
Ultracold atomic gases have been experimentally realized in two-dimensions~\cite{Hadzibabic2006, Bloch2008}, 
and a viable path for implementing IQHE states with ultracold atoms is to employ 
synthetic magnetic fields~\cite{Lin2016,Dal2011,Bloch2012,Goldman2014}. 
The missing ingredient are solenoid-like probes. 
However, the synthetic vector potential of a solenoid can in principle be achieved with vortex 
laser beams non-resonantly interacting with two-level atoms~\cite{Jajtic2018}. 
The setup is sketched in Fig. \ref{fig:experiment}(a).
The advantage of these systems are long coherence times, and the possibility to braid the laser probes.

In a condensed matter setup, we propose to use a 2DEG heterostructure grown on a material with high 
magnetic permeability, $\mu_r\gg 1$. 
Suppose that above the heterostructure one suspends magnetic needles akin to those used in Ref.~\cite{Beche2014}. 
The magnetic needle yields the vector potential of a solenoid (see Fig. \ref{fig:experiment}(b)). 
There is an image of this solenoid in the high-$\mu_r$ material, which together with the real solenoid above, 
provides an effective vector potential of a solenoid probe that pierces the 2DEG. 
The challenging parts of the experiment include creating sufficiently narrow identical needles. The flux needs to be fractional, but not necessarily between $0$ and $\Phi_0$. 
Gauge invariance guarantees that fractional fluxes between any two integers $n\Phi_0$ and $(n+1)\Phi_0$ can be used to synthesize anyons.

\section{Conclusion}

In conclusion, we have demonstrated that synthetic anyons can be implemented upon a noninteracting many-body system by using specially tailored localized (physical) probes which supply the demanded nontrivial topology in the system. The platform used for this demonstration are noninteracting electrons in the IQHE state, and the probes are external solenoids with a magnetic flux that is a fraction of the flux quantum. The same Hamiltonian can describe an ultracold (fermionic) atomic gas in 2D, in a uniform synthetic magnetic field, where the probes are laser beams giving rise to synthetic solenoid gauge potentials. 
The Fermi level is such that only the lowest Landau level states are occupied; 
the localized states which appear at the position of every probe, with energy in the gap, are assumed to be empty. 
We have found the ground state of this system, and demonstrated that it is anyonic in the coordinates of the probes, 
when the flux through solenoids is a fraction of the flux quantum $\alpha \Phi_0$. 
The statistical parameter of synthetic anyons is $\theta = \pi(\alpha-1)$.
We have shown that these synthetic anyons cannot be considered as emergent quasiparticles, and that they cannot be interpreted as Wilczek's charge-flux-tube composites. 
This observation has consequences on the fusion rules of these synthetic anyons, which depend on the microscopic details of the fusion process.

In a future study, it would be interesting to consider forces that act upon the probes. Geometric forces on point fluxes carrying integer quanta of fluxes in quantum Hall fluids were studied in Ref.~\cite{Avron1998}. 
Next, it would be interesting to explore the potential for anyonic physics in a system of solenoids that does not necessarily rely on the quantum Hall effect. For example, one such system might be Aharonov-Bohm billiards~\cite{Berry2010}. 
Finally, it would be interesting to extend the ideas presented here to explore non-Abelian synthetic anyons, and investigate their capacity for topological quantum computing. 

\section{Acknowledgments*}

We acknowledge useful discussions with Silvije Domazet, Mario Basleti\'{c}, Emil Tafra, and Ema Jajti\'{c}. This work was supported by the Croatian Science Foundation Grant No. IP-2016-06-5885 SynthMagIA, and in part by the QuantiXLie Center of Excellence, a project co-financed by the Croatian Government and European Union through the European Regional Development Fund - the Competitiveness and Cohesion Operational Programme (Grant KK.01.1.1.01.0004).

\section{Appendix: Calculation of the ground state}

In this Appendix, we find the spectrum of the single particle ($N_e=1$) Hamiltonian \eqref{Ham} with one solenoid ($N=1$) and vanishing scalar potential ($V = 0$).
\paragraph{Single particle spectrum} 
Putting $\psi = R(r) \exp (i m \varphi)$, with the angular quantum number $m \in \mathbb{Z}$ in the time-independent Schr\" odinger equation $H \psi = E \psi$, and taking the solenoid to be at the origin ($\bm{\eta}_1 = \bm{0}$), we find the radial equation
\[
R'' + \frac{1}{s} R' + \left[\epsilon - (m+\alpha) - \frac{|m+\alpha|^2}{s^2} - \frac{1}{4} s^2 \right]R = 0.
\]
Here $s = r/l_B$, $\epsilon = 2E/(\hbar \omega_B)$ and $(\cdot)' = d/d s(\cdot)$. For $s \gg 1$, the radial equation reduces to
\[
R'' - \frac{1}{4} s^2 R = 0
\]
so that the normalizable asymptotic solution is $R \sim \exp (-s^2 / 4)$. On the other hand, for $s \ll 1$, the radial equation becomes the Cauchy-Euler equation
\[
R'' + \frac{1}{s} R' - \frac{|m+\alpha|^2}{s^2} R = 0
\]
with the solutions $R = s^{\pm |m+\alpha|}$. Let us momentarily concentrate on the solution which is regular at the origin and write $R(s) = s^{|m+\alpha|} \exp(-s^2 / 4) S(s)$. Then the function $S(s)$ is found to satisfy
\begin{align*}
S'' &+ \left(\frac{2 |m+\alpha| + 1}{s} - s \right) S' \\
&+ \left[\epsilon - (m+\alpha) - |m+\alpha| - 1\right]S = 0
\end{align*}
which is the Laguerre differential equation in disguise with the solution $S(s) = L_{n_r}^{|m+ \alpha|} (s^2/2)$, where $n_r = \left[\epsilon - (m+\alpha) - |m+\alpha| - 1\right]/2$ is the radial quantum number. In order not to spoil the normalizable behavior for large $s$, the Laguerre function must reduce to a polynomial which enforces the condition $n_r \in \mathbb{N}_0$, which, in turn, leads to the condition for the allowed energies
\[
\epsilon = 1 + 2 n_r + (m+\alpha) + |m+\alpha|,
\]
which depend on both the radial and angular quantum numbers. The corresponding wavefunctions are
\[
\psi(r,\varphi) = r^{|m+\alpha|} L_{n_r}^{|m+ \alpha|} (r^2/2l_B^2) \exp(-r^2 / 4 l_B^2 + i m \varphi).
\]

\paragraph{Lowest Landau level} Let us now consider the ground state. The minimum energy state is given by the vanishing of the radial quantum number ($n_r = 0$) and the condition $(m+\alpha) + |m+\alpha| = 0$, which is equivalent to $m \leq - \alpha$. With our choice of $\alpha \in \langle 0,1 \rangle$, we have $m \leq -1$. Therefore, the LLL has the energy
\[
\epsilon = 1 \quad \rightarrow \quad E = \frac{1}{2} \hbar \omega_B,
\]
and is infinitely degenerate. The ground state wavefunctions are of the form
\[
\psi_{LLL}(r,\varphi) = r^{|m+\alpha|}\exp(-r^2 / 4 l_B^2 + i m \varphi).
\]
Turning to complex notation, the above wavefunctions are equivalent to
\[
\psi_{LLL}(z) = \frac{\bar{z}}{|z|^\alpha} \bar{z}^m \exp(-|z|^2 / 4 l_B^2), \quad m \in \mathbb{N}_0.
\]
Taking linear combinations of these functions show that the LLL wavefunction can also be written in terms of an arbitrary antiholomorphic function $f(\bar{z})$,
\[
\psi_{LLL}(z) = \frac{\bar{z}}{|z|^\alpha} f(\bar{z}) \exp(-|z|^2 / 4 l_B^2).
\]

\paragraph{Excited state} The first excited state corresponds to the quantum numbers $n_r = 0$ and $m=0$ and belong to the gap between the LLL and the second Landau level. Its energy and wavefunction are
\[
E = \frac{1 + 2 \alpha}{2} \hbar \omega_B, \quad \psi_{LS}(z) = |z|^\alpha \exp(-|z|^2 / 4 l_B^2).
\]
It is easy to see that it is localized around the solenoid probe.

\paragraph{Displaced solenoid} Having obtained the spectrum for the origin-centered solenoid, we can easily generalize our results for arbitrary position of the solenoid $\bm{\eta}$. It suffices to put $z \to z - \eta$ in the above wavefunctions and simultaneously perform a gauge transformation which keeps the vector potential $\bf{A}_0$ unchanged. The needed gauge factor is just
\[
\exp\left[i \frac{q}{2\hbar} (\bf{B} \times \bm{\eta})\cdot \bf{r}\right] = \exp \left( - \frac{\bar{\eta} z - \eta \bar{z}}{4 l_B}\right).
\]
The end result is that the LLL wavefunctions for the displaced solenoid become
\[
\psi_{LLL}(z) =  |z - \eta|^{-\alpha} \conj{z - \eta} f(\bar{z}) \exp(-|z|^2 / 4 l_B^2).
\]
Taking $f(\bar{z}) = \bar{z}^m$, we arrive at the wavefunctions given in Eq.~\eqref{wf}. Using the same approach for the excited state, we arrive at the wavefunction given in Eq.~\eqref{excited}.

\paragraph{Multiple probes} Comparing the solution for the ground state in the case of one displaced solenoid $\psi_{LLL}^{\eta} (z)$ and without solenoids (the usual IQHE ground state) $\psi_{LLL}^{0}(z)$, it is easily seen that the simple relation holds
\[
\psi_{LLL}^{\eta} (z) = \frac{\conj{z - \eta}}{|z - \eta|^{\alpha}} \psi_{LLL}^{0}(z).
\]
This relation motivates the ansatz \eqref{twosol} for the single particle LLL wavefunctions for two (or more) probes which is checked to be correct.

\paragraph{Many-body ground state}

Having obtained the single particle wavefunctions, we can completely fill the lowest Landau level by occupying the $n_r = 0$ state for all possible values of angular quantum number $m$. This amounts to constructing the Slater determinant of single-particle states. Due to the $\bar{z}^m$ term this Slater determinant is of the Vandermonde form and is easily calculated and given by Eq.~\eqref{psiNprobes}.

\paragraph{The spurious divergent solution} Let us now return to the question whether the single particle wavefunction for a single probe centered at the origin can behave singularly as $s^{-|m+\alpha|}$, as we are led to think when solving the radial equation for $s \ll 1$. If that would be the case, then the LLL would have an additional state given by Eq.~\eqref{spurious} which would have to be taken into account when constructing the Slater determinant. This additional state diverges at the origin and corresponds to the angular quantum number $m = 0$. Other divergent solutions exist for $m \neq 0$, but are not even normalizable and therefore are easily excluded to be unphysical. However, the solution Eq.~\eqref{spurious} is normalizable, but can be eliminated on more elementary grounds---it is not the solution to the Cauchy-Euler equation. More specifically, for $\lambda > 0$, one can show by proper regularization of $s^{-\lambda}$ that the following identity holds
\[
\left[\frac{d^2}{d s^2} + \frac{1}{s} \frac{d}{d s} - \frac{\lambda^2}{s^2} \right] s^{-\lambda} = - \frac{2 \lambda}{s^{1 + \lambda}} \delta (s).
\]
This is similar to a more familiar case of 3D Laplacian for which $\bm{\nabla}^2 (1/r) = 0$ everywhere except the origin and so $1/r$ is not a proper harmonic function. Likewise, there are no states in the LLL that show divergent behavior in the vicinity of the probe and one must exclude the state given in Eq.~\eqref{spurious} when constructing the many-body ground state.

\section{Appendix: Calculation of the statistical phase}
\paragraph{Plasma analogy}
To determine the statistics of the probes, we consider a normalized state with $N$ probes given by Eq.~\eqref{psiNprobes}.
Using the plasma analogy, normalization factor $Z(\{ \eta_k \},\{\bar \eta_k\})$ can be interpreted as the partition function of the 2D one-component plasma (electrons) at $\{z_j\}$ at an inverse temperature $\beta=2$, interacting with charged impurities (probes) at $\bm{\eta}_k$~\cite{Laughlin1983, Tong}.
The potential energy for this system is given by 
\begin{equation}
\begin{split}
    U(\{z_j\}) = & \frac{1}{4l_B^2}\sum_{j=1}^{N_e}|z_j|^2 -\sum_{i<j}^{N_e}\log\left(\frac{|z_i-z_j|}{l_B}\right)\\
    & -(1-\alpha)\sum_{j,k}^{N_e,N}\log\left(\frac{|z_j-\eta_k|}{l_B}\right).
    \label{eq:potential_energy}
\end{split}
\end{equation}
In the thermodynamical limit the partition function $Z$ can be obtained by using the saddle-point technique, where the particles are driven into configuration which has the minimum energy \cite{Pasquier2013,Cappelli1993}.
For $N\rightarrow \infty$, the sum over particles becomes a continuous distribution, which equals the electron density. Minimizing the energy and using  $\frac{\partial}{\partial\overline{z}}z^{-1} =\pi\delta^2(z)$, one obtains the density of particles: 
\begin{equation}
    \rho(z)=\frac{1}{2\pi l_B^2}-(1-\alpha)\sum_{k=1}^{N}\delta^2(z-\eta_k).
    \label{eq:density}
\end{equation}
We can recognize two contributions $\rho(z)=\rho_0+\delta\rho(z)$. The first one is constant and corresponds to the density in the case of the $\nu=1$ IQHE, while the second one describes the charge depletion $\Delta q$ at positions of the probes.
In the plasma analogy, an impurity is screened so that its effects cannot be noticed at far distances.
The potential energy and the partition function of the plasma with impurities also include the energy cost between the impurities and the constant background charge, and the Coulomb energy between different impurities. With the additional condition that the corrected partition function $K$ is independent of the positions $\eta_k$~\cite{Tong}, this leads us to the result for the normalization factor
\begin{equation*}
    Z=K\exp{\left(-(1-\alpha)^2\sum_{k<l}^{N}\log\frac{|\eta_k-\eta_l|^2}{l_B^2}+\frac{1-\alpha}{2l_B^2}\sum_{k=1}^{N}|\eta_k|^2\right)}.
\end{equation*}

\paragraph{Berry phase}
In order to find the statistics of the probes, we pick one of the probes, for example $\eta_1$, and move it on a closed path $C$. After traversing the path, the wavefunction
\begin{equation*}
    \psi=\frac{1}{\sqrt{Z}}\chi
\end{equation*}
acquires a phase shift given by the Berry phase 
\begin{equation*}
    e^{i\gamma}=\exp{\left(-i\oint_C \mathcal{A}_{\eta_1}d \eta_1 + \mathcal{A}_{\bar{\eta}_1}d\bar{\eta}_1\right)},
\end{equation*}
where $\mathcal{A}_{\eta_1}$ is the holomorphic and $\mathcal{A}_{\bar{\eta}_1}$ the anti-holomorphic Berry connection:
\begin{equation*}
\begin{split}
    \mathcal{A}_{\eta}(\eta,\bar{\eta})=-\frac{i}{Z}\langle \chi|\frac{\partial}{\partial \eta}|\chi\rangle +\frac{i}{2}\frac{\partial}{\partial\eta}\log{Z},\\
    \mathcal{A}_{\bar{\eta}}(\eta,\bar{\eta})=-\frac{i}{Z} \langle \chi|\frac{\partial}{\partial \bar{\eta}}|\chi\rangle+\frac{i}{2}\frac{\partial}{\partial\bar{\eta}}\log{Z}.
\end{split}
\end{equation*}
The calculation of the Berry phase proceeds as in \cite{Arovas1984}. The braiding phase corresponds to the difference of the Berry phases for closed paths with and without one other probe enclosed by it. 
When $\eta_1$ is taken around the closed path $C$, contributions from the normalization factors, i.e. the partition function $Z$, cancel each other. 
Derivatives of the unnormalized wavefunction $\chi$ are given as
\begin{equation*}
    \frac{\partial \chi}{\partial{\eta_1}}=\frac{\alpha}{2}\chi\sum_{j=1}^{N_e} \frac{1}{z_j-\eta_1},
\end{equation*}
\begin{equation*}
    \frac{\partial \chi}{\partial{\bar{\eta}_1}}=\left(\frac{\alpha-2}{2}\right)\chi \sum_{j=1}^{N_e}\frac{1}{\bar{z}_j-\bar{\eta}_1}.
\end{equation*}
Taking the definition of the charge density 
\begin{equation}
    \rho(z)=\frac{1}{Z} \langle\chi|\sum_{j=1}^{N_e} \delta(z_j-z)|\chi\rangle,
\end{equation}
one obtains
\begin{equation*}
    \gamma=i\frac{\alpha}{2}\int dx dy\oint_{C} d\eta_1\frac{\rho(z)}{z-\eta_1}+ i\frac{\alpha-2}{2}\int dx dy\oint_{C} d\bar{\eta}_1 \frac{\rho(z)}{\bar{z}-\bar{\eta}_1}.
\end{equation*}
If we denote the integral
\begin{equation}
    \mathcal{J}=\oint_{C} d\eta_1\int dx dy \frac{\rho(z)}{z-\eta_1},
\label{eq:J1}
\end{equation}
we have
\begin{equation*}
\gamma=i(\alpha\, \mathrm{Re}\,\mathcal{J}-\bar{\mathcal{J}}).
\end{equation*}
Concerning the contribution of $\rho_0$ in Eq.~(\ref{eq:J1}), if $\eta_1$ is integrated in anticlockwise direction, only values of $z$ inside this loop contribute $-2\pi i$ to the integral. Then we can evaluate the surface integral, where we use the relation between the density and the background magnetic field for the $\nu=1$ IQHE state, $\rho_0=B_0/\Phi_0$.
In order to find the contribution of the second term $\delta\rho(z)$, first we evaluate the surface integral and obtain a non-vanishing contribution from $\eta_k\neq\eta_1$. The contour integral then evaluates to $-2\pi i$ only if $\eta_k$ is inside the closed path of $\eta_1$. 
This leads us to the result
\begin{equation*}
\begin{split}
&\mathcal{J}_{out}=-2\pi i \frac{\Phi_C}{\Phi_0},\\
&\mathcal{J}_{in}=-2\pi i \frac{\Phi_C}{\Phi_0}+2\pi i (1-\alpha),
\end{split}
\end{equation*}
where $\Phi_C$ is the magnetic flux enclosed by the path $C$. Let us denote the mean number of electrons inside the contour as $\langle n\rangle_C$.
Thus, when $\eta_1$ traverses a path where it does not enclose any of the other probes the Berry phase is
\begin{equation*}
\gamma_{out}=2\pi\frac{\Phi_C}{\Phi_0}=2\pi\langle n\rangle_{C,out}.
\end{equation*}
On the other hand, if one other probe is inside the loop, the Berry phase sums up to
\begin{equation*}
\gamma_{in}=2\pi\frac{\Phi_C}{\Phi_0}-2\pi(1-\alpha)=2\pi\langle n\rangle_{C,in}. 
\end{equation*}
The statistical phase is the difference between these two cases
\begin{equation*}
    \gamma_S=2\pi(\langle n\rangle_{C,in}-\langle n\rangle_{C,out})=2\pi(\alpha-1).
\end{equation*}
\paragraph{Fusion}
In this paragraph we find the statistical phase of the fused synthetic anyons. The fusion rule states that the exchange phase $\Gamma_S$ of a particle formed by combining $n$ intentical anyons with exchange phase $\gamma_S$ is $\Gamma_S=n^2\gamma_S$. Suppose that we have two pairs of probes, i.e. $N=4$  solenoid  probes, with flux $\alpha\Phi_0$ in the system located at $\{\eta_1,\eta_2,\eta_3,\eta_4\}$. The wavefunction $\psi$ is given by Eq.~(\ref{psiNprobes}). Two solenoids are paired so that they remain separated by a small constant vector. 
For each pair we use the center-of-mass and relative coordinates
\begin{equation*}
\begin{split}
    X_c=\frac{\eta_1+\eta_2}{2},\quad X_r=\frac{\eta_1-\eta_2}{2};\\
    Y_c=\frac{\eta_3+\eta_4}{2},\quad Y_r=\frac{\eta_3-\eta_4}{2}.
\end{split}
\end{equation*}
If we encircle the first pair of solenoids at $X_c$ along a circle of radius $R$ around the second pair at $Y_c$, which is held static, this process can be described as
\begin{equation*}
    X_c'(\theta)=Y_c+ Re^{i\theta}=Y_c+\lambda,
\end{equation*}
where $\lambda$ is a complex coordinate which moves around the closed path $C$, a circle of radius $R$. 
Then the coordinates of the first pair are moved according to
\begin{equation*}
    \eta_1'=Y_c+X_r+\lambda, \quad \eta_2'=Y_c-X_r+\lambda.
\end{equation*}
The Berry phase acquired in this process is given by
\begin{equation*}
\Gamma=i\int_{0}^{2\pi} d\theta \langle \psi |\frac{\partial}{\partial \theta}\psi\rangle.
\end{equation*}
Since the normalization factor of the wavefunction is single-valued in $\theta$, it does not contribute to the Berry phase for a closed path. 
Taking into account the expression for the charge density and the result
\begin{equation*}
    \frac{d\eta_{1,2}'}{d\theta}=\frac{d\lambda}{d\theta},
\end{equation*}
we obtain
\begin{equation*}
\begin{split}
    \Gamma =&i\frac{\alpha}{2}\oint_{C} d\lambda\int dx dy\left[ \frac{\rho(z)}{z-\eta_1'}+\frac{\rho(z)}{z-\eta_2'}\right]\\
     &+i\frac{\alpha-2}{2}\oint_{C} d\bar{\lambda}\int dx dy\left[\frac{\rho(z)}{\bar{z}-\bar{\eta}_1'}+\frac{\rho(z)}{\bar{z}-\bar{\eta}_2'}\right].
\end{split}
\end{equation*}
Denoting
\begin{equation}
    \mathcal{J}=\oint_C d\lambda \int dx dy \left[ \frac{\rho(z)}{z-\eta_1'}+\frac{\rho(z)}{z-\eta_2'}\right], 
\label{eq:J2}   
\end{equation}
the Berry phase is then 
\begin{equation*}
    \Gamma=i(\alpha\, \text{Re}\, {\mathcal{J}}-\bar{\mathcal{J}}).
\end{equation*}
Regarding the contribution of $\rho_0$ in Eq.~(\ref{eq:J2}), when $\lambda$ is integrated anticlockwise, only values of $z$ inside this path contribute $-4\pi i$ to the integral. As before, one calculates the surface integral, by using the relationship between the density and the magnetic flux in the IQHE state. 
Concerning the contribution of $\delta \rho (z)$, first the surface integral is evaluated. This gives us the result 
\begin{equation*}
\begin{split}
    \mathcal{J}= & -4\pi i \frac{\Phi_C}{\Phi_0}-(1-\alpha)\oint_C d\lambda\Bigg[ \frac{1}{\eta_3-\eta_1'}+\frac{1}{\eta_4-\eta_1'}\\
    & +\frac{1}{\eta_3-\eta_2'}+\frac{1}{\eta_4-\eta_2'}\Bigg].
\end{split}    
\end{equation*}
Evaluating the contour integral, we obtain
\begin{equation*}
    \mathcal{J}=-4\pi i \frac{\Phi_C}{\Phi_0}+8\pi i (1-\alpha),
\end{equation*}
and, finally, the Berry phase is 
\begin{equation}
    \Gamma=4\pi\frac{\Phi_C}{\Phi_0} + 8\pi (\alpha-1).
    \label{eq:fusion_statistical_phase}
\end{equation}
In the second term of the right-hand side of Eq.~(\ref{eq:fusion_statistical_phase}), we can recognize the Aharonov-Bohm phase and the statistical phase $\Gamma_S = 2^2 \times 2 \pi (\alpha-1)$, which confirms the fusion rule for anyons. 


\end{document}